\documentstyle[aps,prl,twocolumn]{revtex}

\newcommand{\p}[2]{\frac{\partial #1}{\partial #2}}
\newcommand{\vx}{{\bf r}}
\newcommand{\vu}{{\bf u}}
\newcommand{\vv}{{\bf v}}
\newcommand{\vw}{{\bf w}}
\newcommand{\vp}{{\bf p}}
\newcommand{\vb}{{\bf B}}
\newcommand{\ve}{{\bf E}}
\newcommand{\be}{\begin{equation}}
\newcommand{\ee}{\end{equation}}
\newcommand{\ba}{\begin{eqnarray}}
\newcommand{\ea}{\end{eqnarray}}

\begin{document}
\draft
\twocolumn[\hsize\textwidth\columnwidth\hsize\csname@twocolumnfalse\endcsname 

\title{An analytic solution of the Boltzmann equation in the
presence of self-generated magnetic fields in astrophysical plasmas}

\author{L.H. Li}
\address{Purple Mountain Observatory, Academia Sinica,
           Nanjing 210008, People's Republic of China}

\maketitle
\begin{abstract}
Through relating a self-generated magnetic field to the
regular motion of free electrons that is responsible for the
magnetic field generation in astrophysical plasmas, we solve the 
Boltzmann kinetic equation in the presence of the self-generated
magnetic fields to obtain a steady-state, collisional invariant analytic
solution of the equation.
\end{abstract}
\pacs{PACS numbers: 52.25.Dg, 95.30.Qd} % \\
%      Keywords: Plasma, Self-generated magnetic
%                fields, Kinetics, Boltzmann equation, 
%                Nonlinearity, Collective behavior\\
%\\
%{\bf Phone:}   086-025-3307609\\
%{\bf Fax:}     086-025-3301459 \\
%{\bf Email:}   linghuai@public1.ptt.js.cn}

] \renewcommand{\thefootnote}{\arabic{footnote}} \setcounter{footnote}{0}

Most astrophysical plasmas are weakly inhomogeneous nonideal gases
consisting of a large number of free charged particles that are closely
coupled together by electrostatic and electromagnetic interactions. These
interactions can propagate in the form of electrostatic waves such as
high-frequency Langmuir waves, low-frequency ion-sound waves, and
electromagnetic waves such as high- and low-frequency transverse plasma
waves\cite{ABR-84}. The so-called self-generated magnetic fields are the
low-frequency (zero-frequency) transverse plasma waves. The zero frequency
waves are defined to be mirror waves.

Spontaneous excitation of the mirror waves is a well-established
experiment fact\cite{SELF-GM} in laser-produced plasmas and
microwave-plasma systems. In these experiments the free energy sources are
coherent radiations. The underlying mechanisms have been well
understood\cite{SC-77}. In addition to coherent radiations, recently it
has been shown that incoherent radiations such as thermal
radiations\cite{LiZhang-96} and neutrinos\cite{SSBB-97} can also excite
magnetic fields in astrophysical plasmas. The keys are the
nonlinear interactions bwteen plasmas and photons or neutrinos.
Bingham, Dawson, Shukla, Stenflo, Bethe etc\cite{SSBB-97} have shown that
nonlinear coupling between plasmas and neutrinos by means of the weak
interaction is likely to be responsible for the most violent 
astrophysical activities such as supernova explosions. In order to take into
account the self-generated magnetic fields in the standard stellar
evolution codes we first have to solve the Boltzmann kinetic equation for the
free electrons in the presence of the self-generated magnetic fields, as we are
about to do in this letter.

Suppose there be a self-generated magnetic field $\vb_s$ in a weakly
inhomogeneous plasma. Our aim is to analyse how such a magnetic field affects
electron distribution in six-dimensional phase space. For the sake of
simplicity, we assume that the plasma consist of electrons and a single species
of ions, the ion mass $m_i$ be much larger than the electron mass $m_e$. Real
plasmas are collisional, so we should start from the Boltzmann kinetic equation
of electrons\cite{ABR-84,Bernstein-88}
\be
 \p{f_e}{t}+\vv_e\cdot\p{f_e}{\vx}
   +{\bf F}_e\cdot\p{f_e}{\vv_e}
    =\left(\p{f_e}{t}\right)^{ee}_{\mbox{\scriptsize{col}}}
     +\left(\p{f_e}{t}\right)^{ei}_{\mbox{\scriptsize{col}}},
     \label{Boltzmann}
\ee
where ${\bf F}_e=({q_e}/{m_e})(\ve_{\mbox{\scriptsize{tot}}}
  +\vv_e\times\vb_{\mbox{\scriptsize{tot}}})$
is the electromagnetic force, noticing that
$\vb_{\mbox{\scriptsize{tot}}}$ contains $\vb_s$. If we exchange
subscripts e and i in Eq.~(\ref{Boltzmann}), we will obtain the
Boltzmann equation for ions.  
\ba
   \left(\p{f_e}{t}\right)^{ei}_{\mbox{\scriptsize{col}}}
     =\int d\vp_i d\vp'_e d\vp'_i
    W(\vp_e,\vp_i;\vp'_e,\vp'_i) \nonumber \\
    \cdot\delta^{(4)}(p_e+p_i-p'_e-p'_i) \nonumber \\
    \cdot[f_e(\vp'_e)f_i(\vp'_i)
      -f_e(\vp_e)f_i(\vp_i)], \label{col}
\ea
is the collision integral between an electron and an ion, in which $W$ is the
collision probability, $p$ is a (4-dimensional) energy-momentum vector, while
the $\delta$ function stands for the energy-momentum conservation law. If the
subscript e is replaced by i in Eq.~(\ref{col}), we will
obtain the collision integral between two electrons.

In order to solve the kinetic equations we utilize the well-established
two-timescale technique\cite{SC-77}. We first decompose $A_e=\{f_e,\
\vp_e,\ \ve_{\mbox{\scriptsize{tot}}}, \vb_{\mbox{\scriptsize{tot}}}\}$
into the fast varying component $A_f$ and the slowly varying
component $A_s$, noticing that the variation time scale is similar
to $\tau_e=2\pi/\omega_{pe}$ for $A_f$,
$\tau_i=2\pi/\omega_{\mbox{\scriptsize{pe(i)}}}$ for $A_s$, where 
$\omega_{\mbox{\scriptsize{pe(i)}}}=(4\pi n_{\mbox{\scriptsize{e(i)}}}
e^2/m_{\mbox{\scriptsize{e(i)}}})^{1/2}$ is the electron (ion) plasma
frequency. Since $\tau_i\gg \tau_e$, $\int_0^{\tau_i}A_f(t) dt=0$. Using such
a decomposion in the Boltzmann equation and taking the time average over
$\tau_i$, we will then obtain
\be
 \p{f_s}{t}+\vv_e\cdot\p{f_s}{\vx}
   +{\bf F}_s\cdot\p{f_s}{\vv_e}
    =\left(\p{f_s}{t}\right)^{ee}_{\mbox{\scriptsize{col}}}
    +\left(\p{f_s}{t}\right)^{ei}_{\mbox{\scriptsize{col}}},
     \label{Boltzm-s}
\ee
where ${\bf F}_s=({q_e}/{m_e})\vv_e\times\vb_s$ in that
$\ve_s=0$ is assumed because of quasineutrality in the slowly varying
time scale. We have neglected those terms that contain $f_f$ since the
condition $f_f\ll f_s$ holds good. Ion distribution function has only slowly
varying component since an ion is much heavier than an electron. So we
have
\be
 \p{f_i}{t}+\vv_i\cdot\p{f_i}{\vx}
   +{\bf F}_i\cdot\p{f_i}{\vv_i}
    =\left(\p{f_i}{t}\right)^{ii}_{\mbox{\scriptsize{col}}}
    +\left(\p{f_i}{t}\right)^{ie}_{\mbox{\scriptsize{col}}},
     \label{Boltzm-i}
\ee
where ${\bf F}_i=({q_i}/{m_i})\vv_i\times\vb_s$. $f_s$ and $f_i$ couple
together through collision integrals.

Generally, particle microscopic velocity $\vv$ can be decomposed into 
a regular motion component $\vu$ and a random (thermal) motion component
$\vw$\cite{kinetic-1,SC-77},
\be
  \vv=\vu+\tilde{w}.
\ee
In order to emphasize randomness of $\tilde{w}$ and distinguish it from the
regular vector $\vu$, we write $\tilde{w}$ explicitly in terms of a random
phase angle $\phi$ and a usual vector $\vw$ as follows
\be
  \tilde{w}=\case{1}{2}\{\vw \exp(-i\phi)+c.c.\}, \label{phi}
\ee
where c.c. represents the complex conjugate of the first term. Obviously,
$\tilde{w}^2=\vw^2$. If $\vb_s=0$, the regular motion velocity components
$\vu_e$ and $\vu_i$ of electrons and ions associated with $\vb_s$ vanish.

When $\vb_s$ is absent, ${\bf F}_s={\bf F}_i=0$ and we thus can apply
Boltzmann's H-theorem\cite{kinetic-1}. According to this thoerem, we know that
the steady-state solutions of Eqs.~(\ref{Boltzm-s}) and (\ref{Boltzm-i})
satisfy
\ba
  f_s(\vx,\vp_1)f_s(\vx,\vp_2)=f_s(\vx,\vp'_1)f_s(\vx,\vp'_2), \nonumber\\
  f_i(\vx,\vp_1)f_i(\vx,\vp_2)=f_i(\vx,\vp'_1)f_i(\vx,\vp'_2), \nonumber\\
  f_s(\vx,\vp_e)f_i(\vx,\vp_i)=f_s(\vx,\vp'_e)f_i(\vx,\vp'_i), \label{detail}
\ea
which are the so-called condition of detailed balance. The first two have
simplest general solutions as such
\ba
  \ln f_s(\vx,\vp_e) &=& \alpha_e(\vx) + \beta_e(\vx) \case{1}{2}m_e[\vw^2
      + \vu_e^2(\vx)], \nonumber\\
  \ln f_i(\vx,\vp_i) &=& \alpha_i(\vx) + \beta_i(\vx) \case{1}{2}m_i[\vw^2
      + \vu_i^2(\vx)], \label{sol-1}
\ea
since they are simplest general scalar functions made of $\tilde{w}$ and $\vu$.
The cross term $\vu\cdot\tilde{w}$ is meaningless and should be
absent because of the randomness of $\tilde{w}$. The combination coefficient
functions $\alpha$ and $\beta$ depend upon macroscopic thermodynamic variables
of the system, particle number density $n(\vx)$ and temperature $T(\vx)$,
defined by
\ba
  n_e(\vx) &=& \frac{N_e}{V} \int f_s(\vx,\vp_e) d\vp_e, \nonumber\\
  T_e(\vx) &=& \case{2}{3} \frac{1}{n_ek_B}\frac{N_e}{V}\int 
    \case{1}{2}m_e\vw^2_ef_s(\vx,\vp_e) d\vp_e, \label{def-1}
\ea
and the same definitions for ions, where $k_B$ is the Boltzmann constant,
$N_e$ the total electron number, $V$ the total volume of the system.
Substituting Eq.~(\ref{sol-1}) into these definitions, using the equality
$d\vp=md\vw$ and the integration formula $\int_0^\infty
x^{2n}\exp(-x^2)dx=(n-\case{1}{2})(n-\case{3}{2})\cdots 
\case{1}{2}\case{1}{2}\sqrt{\pi}$, we obtain
\ba
  \alpha_e &=& \frac{m_e\vu^2_e}{2k_BT_e}+\ln (\frac{n_eV}{N_e})
      -\case{3}{2}\ln(2\pi m_e k_BT_e), \nonumber\\
  \beta_e &=& -\frac{1}{k_B T_e}, \label{alpha}
\ea
and the same expressions for ions. Substituting Eq.~(\ref{sol-1}) with
(\ref{alpha}) into the last equation of Eq.~(\ref{detail}), we find that it
requires
\be
  T_e=T_i=T, \label{temp}
\ee
but imposes no constraint on regular motion velocity components $\vu_e$ and
$\vu_i$.

If a single-component system is in global equilibrium, its particle number
denisty $n=N/V$, regular motion velocity $\vu$ and temperature $T$ do not
depend upon $\vx$. Consequently, we can make such a Lorentz boost that
$\vu=0$ and $\alpha=-\case{3}{2}\ln(2\pi m k_BT)$. In this special case, we
reproduce the well-known Maxwell distribution
\be
  f(\vp)=(2\pi m k_BT)^{-3/2}\exp\left(-\frac{\vp^2}{2mk_BT}\right),
\ee
where $\vw=\vp/m$ is the random thermal motion velocity with respect to the
comoving reference system with the system. In a two-component system, if
$\vu_e\ne\vu_i$, even if the system is in global equilibrium, we cannot find
out a unique comoving reference system so that it comoves with both components
simultaneously. Of cause, if the system is in only local equilibrium, there
is no a unique comoving reference system, either. Therefore, in local
equilibrium described by Eqs.~(\ref{sol-1}), (\ref{alpha}) and (\ref{temp}), if
we retain $\alpha=-\case{3}{2}\ln(2\pi m k_BT)$, the first of Eq.~(\ref{alpha})
reduces to
\be
  n_e(\vx)=\frac{N_e}{V}\exp\left(-\frac{m_e\vu_e^2}{2k_BT} \right),
\ee
and the same expression for ions, which are the natural extension from global
equilibrium to local equilibrium.

When $\vb_s$ is present, we use
\ba
  f_s(\vx,\vp) &=& (2\pi m_e k_BT)^{-3/2}\exp
    \left[-\frac{m_e(\vu_e^2+\vw^2)}{2k_BT}\right], \nonumber\\
  f_i(\vx,\vp) &=& (2\pi m_i k_BT)^{-3/2}\exp
    \left[-\frac{m_i(\vu_i^2+\vw^2)}{2k_BT}\right] \label{sol-2}
\ea
as the try solution of Eqs.~(\ref{Boltzm-s}) and (\ref{Boltzm-i}). Substituting
Eq.~(\ref{sol-2}) into Eqs.~(\ref{Boltzm-s}) and (\ref{Boltzm-i}), using the
explicit expressions of collision integrals, Eq.~(\ref{col}),
we find all collision integrals vanish due to the condition of detailed
balance (\ref{detail}), the terms like ${\bf F}\cdot\partial f/\partial\vv$ vanish since
$(\tilde{w}\times\vb_s)\cdot\tilde{w}=0$ and $(\vu\times\vb_s)\cdot\tilde{w}=0$
(randomness), and $\partial{f}/\partial{t}=0$. Consequently,
Eq.~(\ref{Boltzm-s}) reduces to 
\be
   \nabla\ln(E^{\mbox{\scriptsize{Reg}}}_e/k_BT)=0,
\ee
where $E^{\mbox{\scriptsize{Reg}}}_e=\case{1}{2}m_e\vu_e^2$ is the regular
motion energy per electron. Solving it for $E^{\mbox{\scriptsize{Reg}}}_e$, we
obtain
\be
  E^{\mbox{\scriptsize{Reg}}}_e =\theta k_BT, \label{regular-e1}
\ee
where $\theta$ is a dimensionless integration constant. It does not
explicitly depend upon spatial coordinates but depends upon the magnetic
field $\vb_s$, the plasma temperature $T$, and the number density
of free electrons $n_e$.

In order to relate $\theta$ to those quantities, we need expressing
$E_e^{\mbox{\scriptsize{Reg}}}$ in terms of $\vb_s$ first. In fact, it has been
shown\cite{SC-77} that the self-generated magnetic field is due to a
slowly-varying solenoidal current given by
\be
  {\bf j}_s=-i\frac{e\omega_{\mbox{\scriptsize{pe}}}^2}{16\pi m_e \omega_{0}^3}
    \nabla\times(\ve\times\ve^*), \label{current}
\ee
where $\ve$ is the slowly varying complex envelope of the total
high-frequency electric field $\tilde{\ve}$:
\be
  \tilde{\ve}=\case{1}{2}\{\ve(\vx,t) e^{-i\omega_0 t}
     +\ve^*(\vx,t)e^{i\omega_0 t} \}. \label{high}
\ee
Inasmuch as it is the resonance between the plasma and radiation
fields that is responsible for the magnetic field excitation, $\omega_0\sim
\omega_{pe}$, the high-frequency waves can be considered to be
quasi-monochromatic waves. So we set $\omega_0=\omega_{pe}$ hereafter. 
The validity condition\cite{SC-77} for Eq.~(\ref{current}) is 
\be
 k/k_{\mbox{\scriptsize{De}}}\ll 1 \mbox{ and } 
    (|\ve|^2/8\pi n_e T_e)(k/k_{\mbox{\scriptsize{De}}})\ll 1, \label{valid}
\ee
where $k_{\mbox{\scriptsize{De}}}=(4\pi n_e e^2/T_e)^{1/2}$ is 
the electronic Debye wave number, $k$ is the characteristic wave number.

The Maxwell equation for the slowly varying electromagnetic field
reduces to
$
  \nabla\times\vb_s=4\pi {\bf j}_s
$
in the steady\-state limit. From this equation and Eq.(\ref{current})
we obtain
\be
  \vb_s=-i(e/4 m_e\omega_{pe})
    \ve\times\ve^*, \label{magnet}
\ee
which shows that the magnetic field self-generation in the
plasma is a kind of nonlinear behaviours. If Eq.~(\ref{valid})
is satisfied, we can neglect the convective term in the momentum
balance equation for an 
electron-fluid element. Using this approximation and Eq. (\ref{high}),
we relate the total fast varying electric field $\tilde{\ve}$ to the
fast varying electron-fluid element velocity $\vu_e$,
$
  \vu'_e(\vx,t)=-i(e/{m_e\omega_0})\ve(\vx,t),
$
where $\vu'_e$ is the complex envelope of $\vu_e$, as $\ve$ is the
complex envelope of $\tilde{\ve}$. Using this relation and
defining $\vu'_e=\vu_r+i\vu_i$, Eq.~(\ref{magnet}) can be rewritten as
follows
\be
  \vb_s=-(m_e\omega_{pe}/2e)
     \vu_r\times\vu_i.
\ee

If we measure $\vb_s$ over a much larger time scale $\tau$ than its
correlation time scale $\tau_b$, $\vb_s$ can be considered to be
stochastic. The mean magnetic pressure $P_B$ over $\tau\gg \tau_b$
can be expressed by
\be
  P_B\equiv<|\vb_s|^2/8\pi>_\tau
     =(m^2_e\omega_{pe}^2/{64\pi e^2})
      |\vu_r|^2|\vu_i|^2.
\ee
Assuming $u_r=u_i$ without loss of generality, this equation leads to
\be
  E_e^{\mbox{\scriptsize{Reg}}} = \case{1}{2} m_e<|\vu_e|^2>_\tau
               = (4P_B m_e/n_e)^{1/2}, \label{regular-e2}
\ee
which expresses $E_e^{\mbox{\scriptsize{Reg}}}$ in terms of $P_B$ or
$\vb_s$. Comparing Eq.~(\ref{regular-e1}) with Eq.~(\ref{regular-e2}),
we obtain 
\be
  \theta=\sqrt{2}\frac{\Omega_{ce}}{\omega_{pe}}\frac{m_ec^2}{k_BT},
    \label{theta}
\ee
which expresses the integration constant $\theta$ in terms of the
self-generated magnetic field $B_s$, the plasma temperature $T$ and the number
density of free electrons $n_e$, where $\Omega_{ce}=e\overline{B}_s/m_e$
is the electronic cyclotron frequency with $\overline{B}_s\equiv (8\pi
P_B)^{1/2}$. $\theta$ can be called as magnetic 
parameter since it is determined by the magnetic field $B_s$.
From Equations~(\ref{magnet}) and (\ref{valid}) it can be seen that
\be
  0 < \theta \ll k_{\mbox{\scriptsize{De}}}/k,
\ee
where $L_b\equiv 2\pi/k$ stands for the characteristic correlation length of
the self-generated magnetic field $\vb_s$. If $L_b$ is much larger than the
electronic Debye length 
$r_{\mbox{\scriptsize{De}}}=1/k_{\mbox{\scriptsize{De}}}$ 
as it should be, $\theta$ can be much larger than unity.

For example, using our simulation result\cite{LiZhang-96}
$\overline{B}_s=6\times10^8$ Gauss at the solar center with
$n_e=6.26\times10^{25}$ cm$^{-3}$, $T_c=1.36$ KeV, we can estimate $\theta_c$ 
by means of Eq.~(\ref{theta}) to be $\theta_c=12.5$. Such large a $\theta$ does
not ruin the theory. In fact, numerical simulations\cite{LiZhang-96} show
$L_b\ge 10^{-4}$ cm. Using $T_c$ and $n_e$ we can estimate 
$r_{\mbox{\scriptsize{De}}}=3\times 10^{-9}$~cm. We thus know 
$k/k_{\mbox{\scriptsize{De}}}
  =2\pi r_{\mbox{\scriptsize{De}}}/L_b\le 2\times 10^{-4}$.
Using Eq.~(\ref{magnet}) and $\theta_c$ we can estimate 
$(|\ve|^2/8\pi n_e T_e)(k/k_{\mbox{\scriptsize{De}}})<3\times 10^{-3}$, which
shows that the validity condition (\ref{valid}) of the theory holds good.

Using the magnetic parameter $\theta$, the electron distribution in the
presence of the self-generated magnetic field can be expressed as follows
\be
  f_e(\vx,\vp) = f_{e0}\exp(-\theta),    \label{sts-e}
\ee
where $f_{e0}=(2\pi m_ek_BT)^{3/2}\exp(-m_e\vw^2/2k_BT)$ is the local
maxwellian in that $\vw$ is the random thermal motion velocity.
$\theta k_BT$ takes a role of chemical potentials, called as quasi-chemical
potential\cite{Bernstein-88}. Eq.~(\ref{sts-e}) with $\theta$ defined by
Eq.~(\ref{theta}) is a steady-state, collisional invariant analytic solution of
the Boltamann equation for free electrons in the presence of self-generated
magnetic fields in astrophysical plasmas that are in local thermodynamic
equilibrium. This is the main result of the paper.

With Eq.~(\ref{sts-e}), the statistically-averaged $^7$Be electron capture
rate\cite{Bahcall-62} in stellar interior is equal to
\be
  \lambda_{e7} = \lambda_{e7}^0\exp(-\theta),
     \label{capture}
\ee
see (3.18) of \cite{SSM} for $\lambda^0_{e7}$. Using $\theta_c
= 12.5$ estimated above, we find the suprresion factor of the $^7$Be
solar neutrino flux is $3.6\times10^{-6}$ in the solar core since 
the magnetic field-related regular motion of electrons will hinder
the electron capture reaction of $^7$Be. Such large a suppression factor is
large enough to explain the $^7$Be solar neutrino missing problem, the root of
the solar neutrino problem\cite{FIT}. This, however, tends to worsen the $^8$B
solar neutrino deficit, which shows that even if there be self-generated
magnetic fields in the solar core we still need matter-enhanced neutrino
oscillations to fit\cite{FIT} solar neutrino experiments. Nevertheless, in this
case, the neutrino mixing parameters such as the neutrino mass and mixing angle
may change. This may reduce the difference of the neutrino masses (mixing
angles) inferred from atmospheric neutrino experiments and solar neutrino
experiments. In order to investigate this important problem quantitatively, we
have to use Eq.~(\ref{sts-e}) to re-calculate the equation of state and
opacities of the stellar plasma with the self-generated magnetic fields taken
into account. We'll present such research results separately.

\acknowledgments

This research belonged to project 19675064 supported by NSFC and was also
supported in part by CAS. The author wants to thank Profs. Stenflo and Shukla
for their constructive suggestions on the paper.


\begin{thebibliography}{99}

\bibitem{ABR-84}
A.F. Alexandrov, L.S. Bogdankevich and A.A. Ruk\-hadze,
  Principles of Plasma Electrodynamics (Springer-Verlag, Berlin, 1984).

\bibitem{SELF-GM}
J.A. Stamper, K. Papadopoulos, R.N. Sudan, S.O. Dean and E.A. McLean,
  \prl 26 (1971) 1012;
J.A. Stamper and B.H. Ripin,
  {\em ibid} 34 (1975) 138;
J.J. Thomson, C.E. Max and K. Estabrook,
  {\em ibid} 35 (1975) 663;
W.F. DiVergilio, A.Y. Wong, H.C. Kim and Y.C. Lee,
  {\em ibid} 38 (1977) 541;
J.A. Stamper, E.A. McLean and B.H. Ripin,
  {\em ibid} 40 (1978) 1177;
A. Raven, O. Willi and P.T. Rumsby,
  {\em ibid} 41 (1978) 554;
S.C. Wilks, W.L. Kruer, M. Tabak and A.B. Langdon,
  \prl 69 (1992) 1383;
R.N. Sudan,
  {\em ibid} 70 (1993) 3075.

\bibitem{SC-77}
B. Bezzerides, D.F. DuBois and D.W. Forslund, \pra 16 (1977) 1678;
B. Bezzerides, D.F. DuBois, D.W. Forslund and E.L. Lindman,
  \prl 38 (1977) 495;
P. Mora and R. Pellat,
  Phys. Lett. 66A (1978) 28;
  Phys. Fluids 22 (1979) 2048;
S.A. Bel'kov and V.N. Tsytovich,
  Soviet Phys. JETP 49 (1979) 656;
V.N. Tsytovich,
  Comm. Plasma Phys. Contr. Fusion 4 (1978) 81;
M. Kono, M.M. \v{S}kori\'c and D. ter Haar,
  J. Plasmas Phys. 26 (1981) 123;
O.M. Gradov and L. Stenflo,
  Phys. Lett. 95A (1983) 233;
L.H. Li, Phys. Fluids B 5, (1993a) 350; (1993b) 1760.

\bibitem{LiZhang-96}
L.H. Li and H.Q. Zhang,
  J. Phys. D: Appl.Phys. 29 (1996) 2217.

\bibitem{SSBB-97}
R. Bingham, J.M. Dawson,J.J. Su and H.A. Bethe,
  Phys.Lett. A193 (1994) 279;
J.T. Mendon\c{c}a, R. Bingham, P.K. Shukla, J.M. Dawson and
  V.N. Tsytovich, Phys. Lett. A209 (1995) 78;
R. Bingham, H.A. Bethe, J.M. Dawson, P.K. Shukla and J.J. Su,
  Phys. Lett. A220 (1996) 107;
T. Chiueh, Astrophys. J. 413 (1993) L35;
P.K. Shukla, R. Bingham, H.A. Bethe, J.M. Dawson and L. Stenflo,
  Physica Scripta 55 (1997) 96;
P.K. Shukla, L. Stenflo, R. Bingham, H.A. Bethe, J.M. Dawson 
  and J.T. Mendon\c{c}a, Phys. Lett. A224 (1997) 239;
P.K. Shukla, L. Stenflo, R. Bingham, H.A. Bethe, J.M. Dawson 
  and J.T. Mendon\c{c}a, Phys. Lett. A226 (1997) 375;
P.K. Shukla, L. Stenflo, R. Bingham, H.A. Bethe, J.M. Dawson 
  and J.T. Mendon\c{c}a, Phys. Lett. A230 (1997) 353;
P.K. Shukla, L. Stenflo, R. Bingham, H.A. Bethe, J.M. Dawson 
  and J.T. Mendon\c{c}a, Phys. Lett. A233 (1997) 181.

\bibitem{Bernstein-88}
J. Bernstein, 
  Kinetic Theory in the Expanding Universe (Cambridge University
  Press, Cambridge, 1988).


\bibitem{kinetic-1}
H.J. Kreuzer,
  Nonequilibrium Thermodynamics and its Statistical Foundations
  (Clarendon Press, Oxford, 1981);
D.C. Montgomery and D.A. Tidman,
  Plasma Kinetic Theory (McGraw Hill, New York, 1964).

\bibitem{LiCPZ-97}
L.H. Li, Q.L. Cheng, Q.H. Peng and H.Q. Zhang, \prd 56 (1997) 8082.

\bibitem{Bahcall-62}
J.N. Bahcall, Phys. Rev. 128 (1962) 1297.

\bibitem{SSM}
J.N. Bahcall, Neutrino Astrophysics (Cambridge University
  Press, Cambridge, England, 1989)

\bibitem{FIT}
N. Hata and P. Langacker,
  \prd 56, No.10 (1997);
   52 (1995) 420; 
K. Heeger and R.G.H. Robertson, 
  \prl 77 (1996) 3720;
S. Parke,
  {\em ibid} 74 (1995) 839;
J. Bahcall,
  Phys. Lett B 338 (1994) 276.

\end{thebibliography}
\end{document}